\documentclass{webofc}
\usepackage[varg]{txfonts}   

\usepackage{relsize}
\newcommand{\babar}{\mbox{\slshape B\kern -0.1em{\smaller A}\kern-0.1em B\kern-0.1em{\smaller A\kern -0.2em R }}}

\begin{document}

\title{$XYZ$ at Belle}

\author{
  \firstname{Jens S\"oren} \lastname{Lange}\inst{1}\fnsep\thanks{\email{soeren.lange@exp2.physik.uni-giessen.de}}
  on behalf of the Belle collaboration
}

\institute{Justus-Liebig-Universit\"at Giessen, II.\ Physikalisches Institut, 35392 Giessen, Germany}

\abstract{
  Recent results of exotic states with heavy quarks, denoted as $XYZ$ states, are presented.
  The results include searches for the $Y$(4260) in $B$ meson decays, a spin partner of $Y$(4630), 
  and the strange Pentaquark $P_s^+$. 
  In addition, the measurements of the absolute branching fraction of $X$(3872)
  and search for double $Z_c$ production is discussed. 
}

\maketitle

\section{Introduction}

\label{intro}

In recent years, numerous narrow states with heavy quarks ($c$,$b$) have been observed,
which do not fit the expected properties for charmonium or bottomonium states
from the Cornell potential \cite{potential_1978}. 
For detailed information about the states, their discovery and their properties, 
we refer to the recent reviews \cite{reviews}.  
Throughout this paper below, charge conjugated states and decay modes are implied. 

\section{$Y$(4260) in $B$ meson decays}

\label{c_y4260_in_B_decays}

The vector state $Y$(4260) has been observed in the ISR (initial state radiation) process
$e^+$$e^-$$\rightarrow$$\gamma$$Y$(4260)$\rightarrow$$J$/$\psi$$\pi^+$$\pi^-$
by BaBar \cite{y4260_babar}, Belle \cite{y4260_belle}, CLEO--c \cite{y4260_cleo-c}
and in direct production in $e^+$$e^-$$\rightarrow$$Y$(4260) at BESIII \cite{y4260_bes3}.
Up to date, in total four states $Y$(4008), $Y$(4260), $Y$(4350) and $Y$(4660) have been observed
by several experiments, all of them carrying the quantum numbers $J^{PC}$=1$^{--}$. 
In addition, there are the well known conventional vector charmonium states
$\psi$(4040),  $\psi$(4140), $\psi$(4400) \cite{pdg} in the same mass regime, 
leading to a overpopulation of states and thus pointing to the exotic nature of the $Y$ states. 

In literature, the $Y$(4260) has been discussed as a $D$$\overline{D}_1$(2460) molecule and as a $P$-wave tetraquark \cite{reviews}. 
In particular, the decay to $D^{(*)}$$\overline{D}^{(*)}$ has not been seen yet \cite{pdg}, although the phasespace is large.
This observation can be explained in the context of a $[$$J$/$\psi$$f_0(980)$$]$ hadrocharmonium state 
(with dominant decays to ''core" and ''cloud'' components) or of a $[$($c$$\overline{c}$)$_8g$$]$ hybrid
(with open charm decays being blocked by the gluon fluxtube) \cite{reviews}. 

Searches for the $Y$(4260) in $B$ meson decays have started more than a decade ago
by \babar in the decay
$B^+$$\rightarrow$$K^+$$Y$(4260) with $Y$(4260)$\rightarrow$$J$/$\psi$$\pi^+$$\pi^-$.
In a data set of 211~fb$^{-1}$, 128$\pm$42 signal events were observed \cite{y4260_in_b_decays_babar},
corresponding to a significance of 3.1$\sigma$, and being translated to a range of a branching fraction
of 1.2$<$${\cal B}$$<$2.9$\times$10$^{-5}$. 

Recently, the D0 experiment reported the surprising observation with 167$\pm$41 signal events (4.3$\sigma$ significance) 
in a data set of 10.4 fb$^{-1}$ \cite{y4260_in_b_decays_D0}. However, different from \babar,
$b$-flavoured hadrons ($B$, $B_s$, $B*$, $\Lambda_b$, ...) were selected inclusively instead of $B$ mesons, 
by requiring a $[$$J$/$\psi$$\pi^+$$\pi^-$$]$ system with a secondary vertex. 

\begin{figure}[h]
\centering
\includegraphics[width=\textwidth,clip]{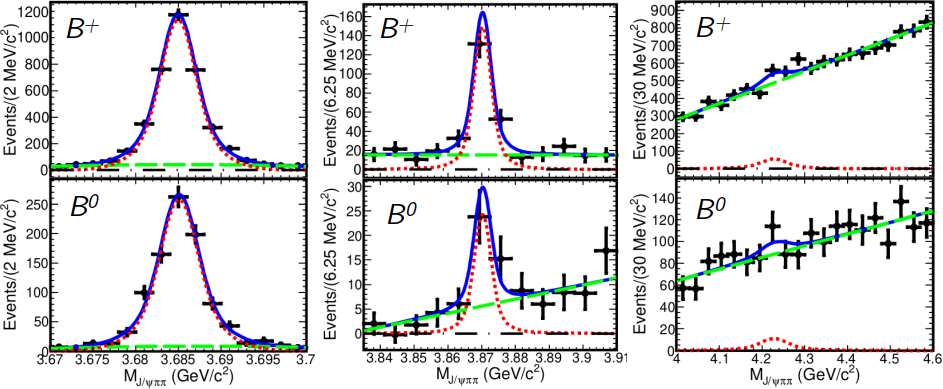}
\caption{Invariant $m$($J$/$\psi$$\pi^+$$\pi^-$) mass distributions
  of $B^+$$\rightarrow$$K^+$$J$/$\psi$$\pi^+$$\pi^-$ decays (top)
  and $B^0$$\rightarrow$$K^0_{s}$$J$/$\psi$$\pi^+$$\pi^-$ decays (bottom)
  at Belle, using the complete recorded data set of 711~$fb^{-1}$. 
  {\it Left:} zoom to mass region of $\psi'$,
  {\it center:} zoom to mass region of $X$(3872), and 
  {\it right:} zoom to mass region of $Y$(4260).\label{fig-1}}
\end{figure}

A search in Belle was performed \cite{y4260_in_b_decays_belle}, using the complete Belle $\Upsilon$(4S) data set of 711~fb$^{-1}$
and corresponding to a factor ~3.3 more data than in case of \babar.
Scaling the signal and background yields would give $S$/$\sqrt{S+B}$=5.7,
which would be a clear observation. 
In addition, neutral B decays with $B^0$$\rightarrow$$K^0_s$$Y$(4260) were added to the analysis.
Fig.~\ref{fig-1} shows the $J$/$\psi$$\pi^+$$\pi^-$ invariant mass distributions for
the three different mass regions around the $\psi'$, the $X$(3872) and the $Y$(4260).
The first two served as control signals. 
For the $Y$(4260), no signal was observed, with the measured branching fraction reaching the 10$^{-6}$ level. 
The derived upper limit of 1.4$\times$10$^{-5}$ for the charged $B$ meson decays is consistent with the \babar measurement.

\section{Search for a spin partner of the $Y$(4630)}

\label{c_y4660_spin_partner}

The $Y$(4630) was discovered in ISR as well and thus also represents a vector state with $J^{PC}$=1$^{--}$.
Along with the $Y$(4660) mentioned above in Sec.~\ref{c_y4260_in_B_decays}, it represents the highest charmed XYZ state with charm observed so far. 
If being a pure $c$$\overline{c}$ state, the radius\footnote{In the Cornell potential, the radius can be approximated
  by the turning point of the wavefunction.}
of can be calculated in the Cornell potential $r$$\simeq$2.2~fm, far beyond string breaking limit
outside the confinement region. 
In addition, it is the only XYZ state observed so far decaying into baryons. 
It was recognized early, that the $Y$(4660), decaying to $J$/$\psi$$\pi^+$$\pi^-$ and already mentioned above, 
and the $Y$(4630), decaying to $\Lambda_c^+$$\Lambda_c^-$, may represent same state \cite{y4660_guo}.
In the hadrocharmonium model, the X(3872) can be interpreted as a $[$$J$/$\psi$$\rho^0$$]$ state, 
the $Y$(4260) as a $[$ $J$/$\psi$ $f_0$(980)$]$ state 
and the $Y$(4630) as a $[$$\psi'$ $f_0$(980)$]$ state. 
The latter one would imply the existence of a $[$$\eta_c$(2S) $f_0$(980)$]$ partner state, called the $Y_\eta$, 
with a mass of 4613$\pm$4 MeV and a width $\simeq$30 MeV, whereas the width is dominated by the $\eta_c$(2S).
With $L$=0, the quantum numbers would be $J^{PC}$=0$^{-+}$, and thus it can not be produced in ISR.
Therefore a search in $B$ meson decays was performed
with the complete Belle $\Upsilon$(4S) data set of 711~fb$^{-1}$ \cite{y4660_spin_partner_belle}. 
No signal was observed yet, leading to an upper limit of
${\cal B}$($B^+$$\rightarrow$$K^+$$Y_\eta$) $\times$ ${\cal B}$($Y_\eta$$\rightarrow$$\Lambda_c^+$$\Lambda_c^-$)$<$2.0$\times$$10^{-4}$.

\section{Absolute branching fraction of the $X$(3872)}

The $X$(3872) has been observed in five different decays \cite{pdg}.
However, all branching fractions are only established as lower limits \cite{pdg},
because the total branching fraction ${\cal B}_{total}$,
being required as normalisation, is not known. 
There is a way to measure ${\cal B}_{total}$ at $B$ factories,
which operate with $e^+$$e^-$ collisions at the $\Upsilon$(4S) resonance at $\sqrt{s}$=10.58~GeV.
The particular situation kinematic situation at the $\Upsilon$(4S) is, 
that the mass of the $\Upsilon$(4S)) coincides very closely with the sum of the masses
of the $B$ meson and the $\overline{B}$ meson, the decay products of the $\Upsilon$(4S). 
Therefore, the two $B$ mesons are almost at rest in the center-of-mass (cms) system.
One of the two $B$ mesons can now be tagged, with a hierarchical full reconstruction
of 1104 hadronic decays \cite{neurobayes}.
The other, remaining $B$ meson could be decay into a $K$ meson and a charmonium or charmonium-like,
exotic state $X_{c\overline{c}}$.
Due to the peculiar kinematic situation, the missing mass $m_{miss}$ of the $K$ meson  
can be utilized for the measurements of total cross sections of any $X_{c\overline{c}}$, and in particular the X(3872). 
The missing mass is calculated as $m_{miss}$($K$)=$\sqrt{p^*_{e^+e^-} - p^*_{tag} - p^*_K)^2}$/$c^2$
with cms momenta of the $e^+$$e^-$ collision system, the tagging side and the $K$ meson, respectively. 
The recently measured value is ${\cal B}_{total}$=1.2$\pm$1.1$\pm$0.1$\times$$10^{-4}$ \cite{belle_x3872_absolute}, 
leading to an upper limit of ${\cal B}_{total}$$<$2.6$\times$$10^{-4}$.
Interestingly, this value almost reaches the value of
the highest known partial product branching fraction \cite{pdg} for 
$B^+$$\rightarrow$$K^+$$X$(3872) with $X$(3872)$\rightarrow$$D^0$$\overline{D}^0$$\pi^0$
of ${\cal B}_{partial}$=(1.0$\pm$0.4)$\times$10$^{-4}$.
Thus, the measurement of ${\cal B}_{total}$ is in close reach for the ongoing data taking at Belle II. 

\section{Double $Z_c$ production}

The charged $Z_c$ states have been observed in $e^+$$e^-$$\rightarrow$$Z_c^+$(3900)$\pi^-$
at BESIII at $\sqrt{s}$=4.26~GeV \cite{z3900_bes3}. 
In addition, the charged $Z_b$ states have been observed in $e^+$$e^-$$\rightarrow$$Z_b^+$$\pi^-$
at Belle at higher $\sqrt{s}$=10.86~GeV in the bottomonium mass regime. 
Thus, it is intuitive to perform a search for $e^+$$e^-$$\rightarrow$$Z_c^+$$Z_c^{(')-}$ pair production
at such higher cms energies of $\sqrt{s}$=10.52~GeV, 10.58~GeV and 10.86~GeV, using the complete Belle data set. 
This search was performed at Belle \cite{zczc_belle}, adding even 102 Mill. $\Upsilon$(1S) and 158 Mill. $\Upsilon$(2S) decays
at $\sqrt{s}$=9.46~GeV and 10.02~GeV, respectively.
$Z_c^+$ $Z_c^{(')-}$ refer to any possible combination of the $Z_c$(3900) and the $Z_c$(4200) states.
No signal was observed, resulting, as an example, in an upper limit on the product branching fraction
$\sigma$$\times$${\cal B}$($Z_c^+$(3900)$\rightarrow$$J$/$\psi$$\pi^+$) of 2.3~fb
for the $Z_c$(3900)$Z_c$(3900) combination. 
The order of magnitude of these upper limits is about a factor of 10$^6$ smaller than 
typical cross sections of $\sigma$($e^+$$e^-$$\rightarrow$$\Upsilon$(5S))=0.3~nb at $\sqrt{s}$=10.86~GeV
and about a factor 10$^4$ smaller than 
$\sigma$($e^+$$e^-$$\rightarrow$Y(4260))$\simeq$65~pb at $\sqrt{s}$=4.26~GeV, 
with $\Upsilon$(5S)$\rightarrow$$Z_b^{(')+}$$\pi^-$ and $Y$(4260)$\rightarrow$$Z_c^+$(3900)$\pi^-$. 

\section{Search for the $P_s^+$ pentaquark}

\begin{figure}[h]
\centering
\includegraphics[width=0.65\textwidth,clip]{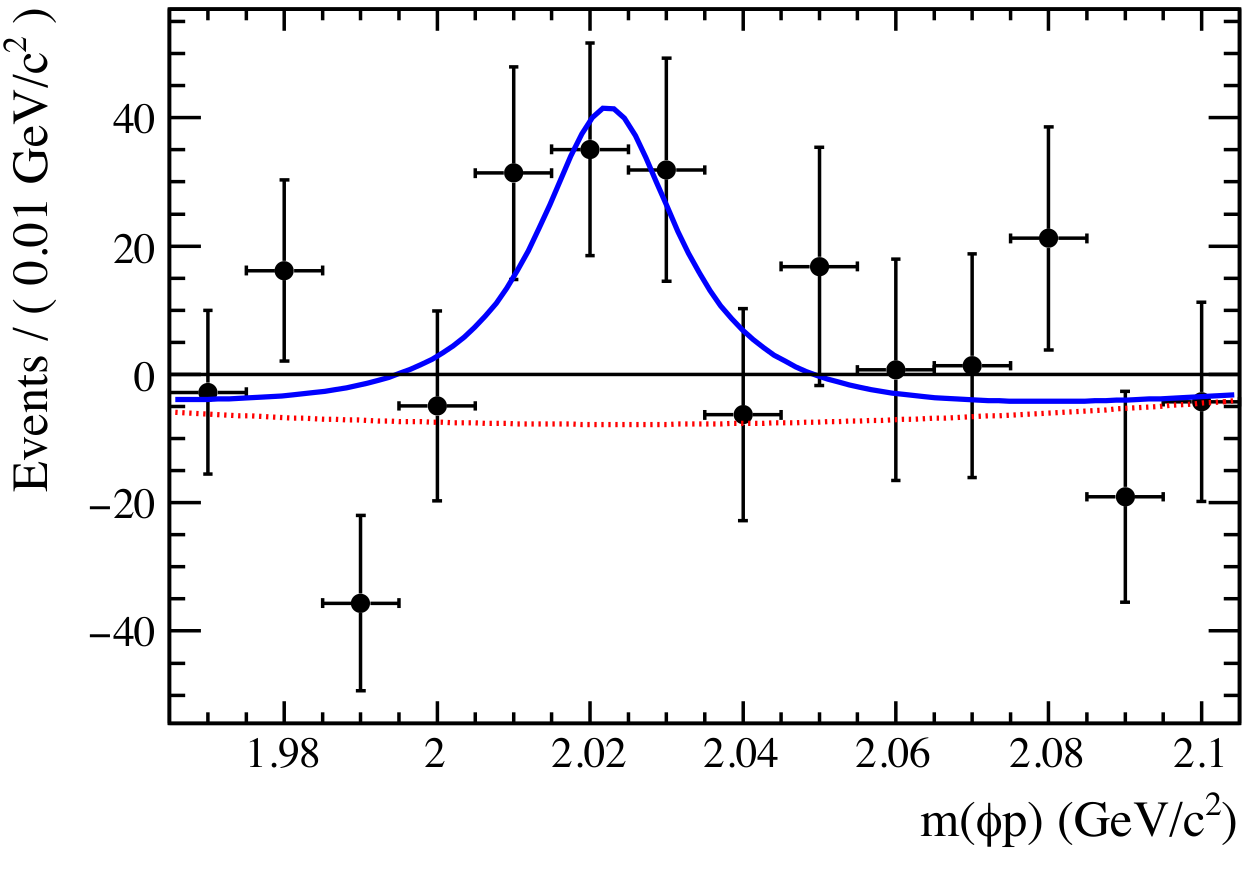}
\caption{Invariant $m$($\phi$$p$) mass distributions in $\Lambda_c$$\rightarrow$$\phi$$p$$\pi^0$ decays
  in the search for the $P_s^+$ pentaquark state by Belle. For details about the background subtraction
  and the fit procedure see \cite{Ps_belle}. The red dotted line shows the phasespace distribution, which after
  background subtraction exhibits a negative fluctuation.\label{fig-2}}
\end{figure}

The charmed pentaquark candidate $P_c^+$$\rightarrow$$J$/$\psi$$p$ has been observed
in $\Lambda_b$ decays at LHCb \cite{Pc_lhcb}. 
For Belle (II), the $\Lambda_b$ is outside the kinematically accessible range.
However, $\Lambda_c$ baryons are produced copiously and provide a suitable data set for the search
for the strange partner of the $P_c^+$, called $P_s^+$, in the decay $P_s^+$$\rightarrow$$\phi$$p$
in $\Lambda_c$$\rightarrow$$[$$\phi$$p$$]$$\pi^0$. 
The search was performed at Belle \cite{Ps_belle}
in a data set of 915 fb$^{-1}$ with total 1.468.435$\pm$4816 $\Lambda_c$ candidates. 
Fig.~\ref{fig-2} shows the background subtracted $m$($\phi$$p$) distribution. 
Each data point corresponds to a fitted yield of a 2D fit.
For details about the background subtraction and the fit procedure see \cite{Ps_belle}. 
No significant signal is observed. 
The fit result of 77.6$\pm$28.1 $P_s$ events 
at a mass of 2.025$\pm$0.005~GeV/$c^2$ and
a width of 0.022$\pm$0.012 GeV/$c^2$ leads
to an upper limit of the product branching fraction of 

\begin{displaymath}
{\cal B} ( \Lambda_c^+ \rightarrow P_s^+ \phi ) \times {\cal B} ( P_s^+ \rightarrow \phi p )
< 8.3 \times 10^{-5}.  
\end{displaymath}

The determined upper limit for $P_s^+$ is still about a factor of 6 higher
than measured branching fraction for the $P_c^+$ at LHCb.

\section{Acknowledgement}

We would like to warmly thank the colleagues of BINP and Novosibirsk State University for their kind hospitality. 
This project was supported by the European Union’s Horizon 2020 research and innovation program 
under grant agreement No.\ 644294 and the German Ministry of Education and Research (BMBF) under grant agreements 05H15RGKBA and 05H19RGKBA.

\newpage

\newpage

\begin{thebibliography}{}


\bibitem{potential_1978}
E.~Eichten, K.~Gottfried, T.~Kinoshita, K.~D.~Lane, T.-M.~Yan, Phys.~Rev.~{\bf D17}, 3090 (1978) 
  
\bibitem{reviews}
A.~Esposito, A.~Pilloni, A.~D.~Polosa, Phys.~Rep.~{\bf 668}, 1 (2016);
R.~F.~Lebed, R.~E.~Mitchell, E.~S.~Swanson, Prog.~Part.~Nucl.~Phys.~{\bf 93}, 143 (2017);
A.~Ali, J.~S.~Lange, S.~Stone, Prog.~Part.~Nucl.~Phys.~{\bf 97}, 123 (2017);
S.~L.~Olsen, T.~Skwarnicki, D.~Zieminska, Rev.~Mod.~Phys.~{\bf 90}, 015003 (2018);
F.–K.~Guo, C.~Hanhart, U.–G.~Mei\ss{}ner, Q.~Wang, Q.~Zhao, B.~S.~Zou, Rev.~Mod.~Phys.~{\bf 90}, 015004 (2018)
  
\bibitem{y4260_babar}
B.~Aubert et al.\ (Babar Collaboration), Phys.\ Rev.\ Lett.\ {\bf 95}, 142001 (2005)

\bibitem{y4260_belle}
C.~Z.~Yuan et al.\ (Belle Collaboration), Phys.\ Rev.\ Lett.\ {\bf 99}, 182004 (2007)

\bibitem{y4260_cleo-c}
Q.~He et al.\ (CLEO-c Collaboration), Phys.\ Rev.\ {\bf D74}, 091104 (2006)

\bibitem{y4260_bes3}
M.~Ablikim et al.\ (BESIII Collaboration), Phys.\ Rev.\ Lett.\ {\bf 118}, 092001 (2017)

\bibitem{pdg}
M.~Tanabashi et al.\ (Particle Data Group), Phys.\ Rev.\ {\bf D98}, 030001 (2018)
  
\bibitem{y4260_in_b_decays_babar}
B.~Aubert et al.\ (BaBar Collaboration), Phys.\ Rev.\ {\bf D73}, 011101 (2006)

\bibitem{y4260_in_b_decays_D0}
V.~M.~Abazov et al.\ (D0 Collaboration), Phys.\ Rev.\ {\bf D98}, 052010 (2018)

\bibitem{y4260_in_b_decays_belle}
R.~Garg, V.~Bhardwaj, J.~B.~Singh, et al. (Belle Collaboration), Phys.~Rev.~{\bf D99}, 071102 (2019)


\bibitem{y4660_guo}
F.-K.~Guo, J.~Haidenbauer, C.~Hanhart, U.-G.~Mei\ss{}ner, Phys.\ Rev.\ {\bf D82}, 094008 (2010)

\bibitem{hadrocharmonium}
S.~Dubynskiy, M.~B.~Voloshin, Phys.\ Lett.\ {\bf B666}, 344 (2008)

\bibitem{y4660_spin_partner_belle}
Y.~B.~Li, C.~P.~Shen, et al. (Belle Collaboration), Eur.\ Phys.\ J.\ {\bf C78}, 252 (2018)


\bibitem{belle_x3872_absolute}
Y. Kato, T. Iijima, et al. (Belle Collaboration), Phys.\ Rev.\ {\bf D97}, 012005 (2018) 

\bibitem{neurobayes}
M.~Feindt, F.~Keller, M.~Kreps, T.~Kuhr, S.~Neubauer, D.~Zander, A.~Zupanc, NIM\ {\bf A654}, 432 (2011)


\bibitem{Pc_lhcb}
R.~Aaij et al.\ (LHCb Collaboration), Phys.\ Rev.\ Lett.\ {\bf 115}, 072001 (2015) 

\bibitem{y2175_babar}
B.~Aubert et al.\ (\babar Collaboration), Phys.\ Rev.\ {\bf D74}, 091103 (2006)

\bibitem{y2175_belle}
C.~P.~Shen et al.\ (Belle Collaboration), Phys.\ Rev.\ {\bf D80}, 031101 (2009)

\bibitem{y2175_besiii}
M.~Ablikim et al.\ (BESIII Collaboration), Phys.\ Rev.\ {\bf D91}, 052017 (2015)

\bibitem{Ps_belle}
B.~Pal, A.~J.~Schwartz, et al.\ (Belle Collaboration), Phys.\ Rev.\ {\bf D96}, 051102 (2017) 


\bibitem{z3900_bes3}
M.~Ablikim at al.\ (BESIII Collaboration), Phys.\ Rev.\ Lett.\ {\bf 110}, 252001 (2013)

\bibitem{zb_belle}
A.~Bondar et al.\ (Belle Collaboration), Phys.\ Rev.\ Lett.\ {\bf 108}, 122001 (2011) 

\bibitem{zczc_belle}
S.~Jia, C.~P.~Shen, C.~Z.~Yuan, et al.\ (Belle Collaboration), Phys.\ Rev.\ {\bf D97}, 112004 (2018)

\end{thebibliography}
\end{document}